%% file: main.tex
\documentclass[10pt, conference, letterpaper]{IEEEtran}
\usepackage{enumitem}
\usepackage{url}
\usepackage{graphicx}
\usepackage{multirow}
\usepackage{comment}
\usepackage{xcolor}
\usepackage{dblfloatfix}
\usepackage{lipsum}
\usepackage{comment}
\usepackage{amsmath,amssymb}
\usepackage{hhline}
\usepackage{tabularx,ragged2e}
\usepackage{lipsum}
\usepackage{ragged2e}
\usepackage[hidelinks]{hyperref}
\usepackage{tikz}
\usepackage{cite}
\usepackage{esvect}
\usepackage{soul}
\usepackage{mathtools}
\usepackage{verbatim}
\newcolumntype{C}{>{\Centering\arraybackslash}X}

\newtheorem{problem}{Problem}

\usepackage{algorithm}
\usepackage{algorithmic}
\usepackage{xspace}

\newcommand{\name}{Kairos\xspace}

\usepackage{calc}
\newlength\myheight
\newlength\mydepth
\settototalheight\myheight{Xygp}
\ExplSyntaxOn
\NewDocumentCommand{\prow}{m}
 {
  \seq_set_split:Nnn \l_tmpa_seq { , } { #1 }
  \seq_use:Nn \l_tmpa_seq { & }
  \\
 }
\ExplSyntaxOff

\begin{document}

 \title{\name: Energy-Efficient Radio Unit Control for O-RAN via Advanced Sleep Modes \vspace{-3mm}\\
{\large As accepted in the 2025 IEEE INFOCOM conference}\vspace{-5mm}
}

\author{\IEEEauthorblockN{J. Xavier Salvat Lozano\IEEEauthorrefmark{1}, Jose A. Ayala-Romero\IEEEauthorrefmark{1}, Andres Garcia-Saavedra\IEEEauthorrefmark{1} and
Xavier Costa-Perez\IEEEauthorrefmark{2}\IEEEauthorrefmark{1}\IEEEauthorrefmark{3}}
\IEEEauthorblockA{NEC Laboratories Europe\IEEEauthorrefmark{1}, i2CAT Foundation\IEEEauthorrefmark{2}, ICREA\IEEEauthorrefmark{3}\\ 
}
\vspace{-6mm}
}

\maketitle

\IEEEoverridecommandlockouts

\IEEEpubid{\begin{minipage}{\textwidth}\ \\[12pt]
\copyright2025 IEEE.  Personal use of this material is permitted.  Permission from IEEE must be obtained for all other uses, in any current or future media, including reprinting/republishing this material for advertising or promotional purposes, creating new collective works, for resale or redistribution to servers or lists, or reuse of any copyrighted component of this work in other works.
\end{minipage}} 

\begin{abstract}
The high energy footprint of 5G base stations, particularly the radio units (RUs), poses a significant environmental and economic challenge. We introduce \name, a novel approach to maximize the energy-saving potential of O-RAN's Advanced Sleep Modes (ASMs). Unlike state-of-the-art solutions, which often rely on complex ASM selection algorithms unsuitable for time-constrained base stations and fail to guarantee stringent QoS demands, Kairos offers a simple yet effective joint ASM selection and radio scheduling policy capable of real-time operation. This policy is then optimized using a data-driven algorithm within an xApp, which enables several key innovations: ($i$) a dimensionality-invariant encoder to handle variable input sizes (e.g., time-varying network slices), ($ii$) distributional critics to accurately model QoS metrics and ensure constraint satisfaction, and ($iii$) a single-actor-multiple-critic architecture to effectively manage multiple constraints.  Through experimental analysis on a commercial RU and trace-driven simulations, we demonstrate \name's potential to achieve energy reductions ranging between 15\% and 72\% while meeting QoS requirements, offering a practical solution for cost- and energy-efficient 5G networks.
\end{abstract}


\input{introduction}

\input{analysis}

\input{mechanism_design}

\input{policy_opt}

\input{perf_eval}
\input{related_work}

\input{conclusions}
\input{ack}

\bibliographystyle{IEEEtran}
\bibliography{references}

\end{document}

%% file: introduction.tex
\section{Introduction}

5G base stations (BS) consume four times more energy than their 4G counterparts~\cite{han2020energy}.  This poses a serious environmental concern: with over 3.5 million 5G BSs deployed worldwide (and growing), the resulting carbon emissions caused by 5G BSs alone are estimated to exceed 100 \emph{megatons} per year~\cite{china_carbon}. This also presents a financial challenge for mobile operators, potentially adding EUR 26.5 billion per year\footnote{Assuming 4.3 kW per 5G BS~\cite{han2020energy} and EUR  0.2008 per kWh of electricity costs for non-households in Europe~\cite{eurostat_energy}. \vspace{4mm}} to electricity bills, exacerbating already high operating expenses. Additionally, the higher power demand may necessitate retrofitting 30\% of power supply systems, incurring an additional cost of approximately USD 2,800 \emph{per site} in capital expenditure~\cite{huawei_energy}.

We make two key observations. First, this four-fold increase in power consumption is primarily attributed to the increased bandwidth, transmission power, and number of antennas that can be employed in 5G macro-cells.  The radio unit (RU) accounts for 90\% of this energy bill~\cite{han2020energy}, which highlights a critical need to curb the RU's energy drain in 5G macro-cells.

\begin{figure}[t!] 
\vspace{-4mm}
\centering
\begin{minipage}{\linewidth}
\centering
\includegraphics[width=\columnwidth]{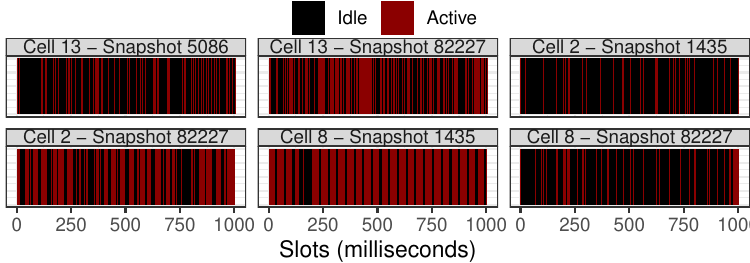}
\end{minipage}
\vspace{-4mm}
\caption{ Short and bursty energy-saving opportunities in production cells. Red/black slots are 1-ms periods with/without transmissions. }
\vspace{-6mm}
\label{fig:motivation:burstiness}
\end{figure}

Second, while very few BSs experience zero traffic over longer time scales (tens of minutes), at shorter time resolutions (milliseconds), most cells are idle most of the time~\cite{8486619}. We confirm this in \S\ref{sec:exp_analysis} through our own measurements on cells from various operators in Madrid, Spain, in 2024. For illustration, Fig.~\ref{fig:motivation:burstiness} shows activity patterns for a few of those cells. In general, as we show in \S\ref{sec:exp_analysis}, the cumulative idle time amounts to over 50\% of the total time in the median cell.

These two observations reveal substantial energy-saving potential through discontinuous transmissions (DTX), a fast switching mechanism that enables deactivating the RU's power amplifier---the most energy-consuming component, as we also show in \S\ref{sec:exp_analysis}---during these short, bursty idle periods~\cite{5956235}. Even a modest 1\% reduction in RU consumption could potentially save over 1,200 GWh annually.

More recently, the O-RAN Alliance introduced specifications for controlling Advanced Sleep Modes (ASMs) in the RU, offering deeper sleep levels and greater energy savings through progressive deactivation of RU components~\cite{oran_asm}. However, deeper sleep levels come at the cost of increased switching overhead. Several research works have since proposed strategies to optimize the use of ASMs, such as~\cite{salem2019traffic, asm_digital_twin}. However, these works often make limiting assumptions about traffic models, neglect the computational complexity and real-time requirements of their solutions~\cite{gpf}, and fail to provide hard QoS guarantees. A comprehensive review of related literature is presented in \S\ref{sec:related}.

To address these issues, we propose a different approach, which we call \name. Instead of relying on complex ASM control algorithms within the BS, which has real-time constraints, we advocate for a simple \emph{joint ASM selection/radio scheduling policy} that is dynamically optimized in \emph{near-real-time} based on context. This optimization is performed within an xApp in the O-RAN Near-Real-Time RAN Intelligent Controller (Near-RT RIC), with more relaxed timing requirements~\cite{garcia2021ran}. This allows us to design a data-driven algorithm with key innovations: ($i$) support for a \emph{varying} number of network slices  over time, with diverse/changing QoS requirements, and ($ii$) inciting joint ASM/radio policies that can minimize energy usage while providing \emph{hard} QoS guarantees.

\name's xApp-based approach has two key advantages: ($i$) it is suitable for deployment in systems with real-world constraints, as the computationally intensive optimization is \newpage \noindent offloaded to the Near-RT RIC, which has more relaxed timing constraints than the BS; and ($ii$) because the intelligence resides in an xApp and is observation-driven, \name can operate with various ASM selection strategies, including those proposed in \cite{salem2019traffic, asm_digital_twin} or others reviewed in \S\ref{sec:related}. However, our results show that, when using \name, complex real-time ASM selection strategies do not yield substantial gains over simpler ones, such as the one we introduced in \S\ref{sec:design:mac}. This is because, as our measurements on a commercial RU in \S\ref{sec:exp_analysis} demonstrate, consolidating radio resources into as few time slots as possible is sufficient for maximizing energy efficiency. Moreover, to provide hard QoS guarantees, the ASM strategy must be directly guided by the configured radio scheduling policy, an aspect that is neglected in the literature.

%% file: analysis.tex
\section{Background \& Analysis}\label{sec:exp_analysis}\label{sec:background}

First, we provide background on 5G and O-RAN (\S\ref{sec:background:primer}), empirically study energy-saving opportunities in real-world networks (\S\ref{sec:analysis:opportunities}), analyze the power consumption profile of massive MIMO 5G radio units (\S\ref{sec:analysis:power}), and detail relevant aspects of Advanced Sleep Modes and O-RAN control (\S\ref{sec:background:asm}).

\subsection{A primer on 5G New Radio (NR) and O-RAN}\label{sec:background:primer}

NR defines the PHY/MAC procedures in 5G. We focus on sub-6GHz bands, which offer up to 100 MHz per band and flexible numerology $\mu = \{0, 1, 2\}$. The fundamental spectrum unit is the resource block (RB), consisting of 12 subcarriers spaced at $15 \cdot 2^\mu$~KHz. Time is divided into 1-ms subframes, each containing $2^\mu$ slots carrying, typically, 14 OFDM symbols of duration $66.7 \cdot 2^{-\mu}$~$\mu$s. In each Transmission Time Interval (TTI), often one slot,  a scheduler assigns one transport block (TB) for each active User Equipment (UE). The TB size depends on the numerology, the buffered data, a RB/symbol scheduling policy, and the modulation and coding scheme (MCS), which depends on the signal-to-noise ratio (SNR).

To break vendor lock-in and enhance flexibility, the O-RAN Alliance has proposed a novel, open, and interoperable architecture for cellular networks with standardized interfaces~\cite{garcia2021ran}. Specifically, O-RAN establishes a 7-2x functional split between  a Distributed Unit (DU)~\cite{garcia2021ran}, which performs higher PHY functions (e.g., forward error correction), a Radio Unit (RU), in charge of lower PHY functions (e.g., FFT). 

An open fronthaul (FH) interface, based on eCPRI, supports this configuration by providing a standardized connection between the RU and the DU. In this interface, a \emph{section} is a structured container that carries specific types of information between the DU and the RU. User Plane (U-Plane) sections carry user data, Control Plane (C-Plane) sections transport control information, and Management Plane (M-Plane) sections carry information used for management. 

Finally, the Near-Real-Time RAN Intelligent Controller (Near-RT RIC) interacts with an O-RAN-compatible DU (O-DU) via the E2 interface, enabling the exchange of fine-grained performance measurements and radio scheduling policies at a $\sim\!100$~ms granularity.

\subsection{Small-timescale energy-saving opportunities}\label{sec:analysis:opportunities}

Since mobile systems are typically provisioned to handle peak traffic demands, individual cells often operate at low average utilization levels~\cite{cloudric}. This inherent underutilization has motivated extensive research aimed at leveraging long-term traffic variations to improve energy efficiency through cell load consolidation and cell sleeping strategies. A comprehensive review of related work is provided in \S\ref{sec:related}.

However, workload patterns within individual cells reveal substantial power-saving sleeping opportunities at shorter timescales. To illustrate this,  using Falcon~\cite{Falkenberg2019a}, we collected millisecond-resolution workload data from diverse cells operated by multiple providers in Madrid, Spain, in May 2024. Consistent with prior research~\cite{cloudric}, the overall load is low (median less than 2~Mb/s) but highly variable (99th percentile around 10~Mb/s). For perspective, a 5G  100-MHz massive MIMO macro-cell can easily reach 1 Gb/s of peak capacity.

\begin{figure}[t!]
 \centering
  \minipage{0.47\columnwidth}
    \includegraphics[width=\columnwidth]{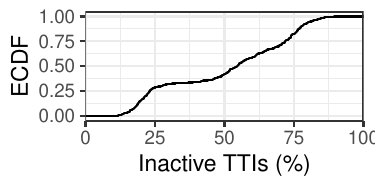}
    \vspace{-7mm}
    \caption{ Share of Idle TTIs in 10-minute periods across cells.}
    \label{fig:motivation:cdf}
 \endminipage{}
 \hfill
 \minipage{0.47\columnwidth}
     \includegraphics[width=\columnwidth]{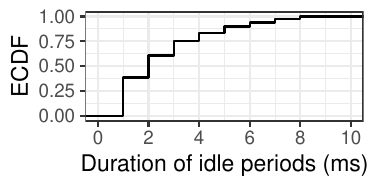}
     \vspace{-7mm}
     \caption{ Duration of idle periods across different cells.}
      \label{fig:motivation:iat}
 \endminipage{}
 \vspace{-6mm}
 \end{figure}

Moreover, we classified each TTI (1 ms in this case) as either "Active" (carrying traffic) or "Idle" (no traffic). Fig.~\ref{fig:motivation:burstiness} shows some activity patterns, \emph{revealing the bursty nature of today's cellular traffic and highlighting potential energy savings through short-term sleep mechanisms.} Indeed, these short-term sleeping opportunities can amount to significant energy savings. Fig.~\ref{fig:motivation:cdf} shows the distribution of the ratio of idle TTIs across all cells, calculated over 10-minute periods: the median inactivity is over 50\%.

However, the duration of such idle periods is highly diverse. To illustrate this, Fig.~\ref{fig:motivation:iat} depicts their distribution across all monitored cells. The median idle period duration is 2 TTIs (28 OFDM symbols), but the 99th percentile reaches 8 TTIs (112 OFDM symbols). \emph{This variability motivates the use of diverse sleep modes that can be implemented at intervals as short as one OFDM symbol or at longer intervals, with energy savings adjusted based on the specific case.} O-RAN refers to these modes as Advanced Sleep Modes (ASM)---see \S\ref{sec:background:asm}.

\subsection{Radio Unit power consumption}\label{sec:analysis:power}

We conducted an experimental analysis on a commercial 32TRX massive MIMO RU compliant with O-RAN 7.2-x FH specifications.\footnote{Details omitted due to Non-Disclosure Agreement with the vendor.} Based on these measurements, we can model the energy consumption of an RU as
\begin{align}\label{eq:ru_power}
    P = P_0 + P_{RF} + P_{BB} + P_{PAM}
\end{align}
where
 \begin{itemize}[noitemsep,topsep=0pt,parsep=0pt,partopsep=0pt,leftmargin=8pt]
    \item $P_{BB}$ models the power consumption of the core digital processor (e.g., an eFPGA) and the other low-PHY processing components (e.g., beamforming) for each antenna chain.
    \item $P_{RF}$ models the power consumption of the analog-to-digital/digital-to-analog converters (ADC/DAC) and the frequency up/downconverters.
    \item $P_0$ models the power consumption of the baseline components in the RU (fabric, Ethernet, PCIe, etc.).
    \item $P_{PAM}$ models the power consumption of the power amplifier module (PAM) at each antenna chain. 
\end{itemize}

Our experiments suggest that $P_0$, $P_{RF}$, $P_{BB}$ remain rather static, independent of the load. However, $P_{PAM}$ is highly dependent on the load and its consumption follows:
\begin{align}
 P_{PAM} = P_{PAM, 0}\cdot P_{PAM, c}(r)\cdot \eta_{PAM} \cdot n_{ant}
\end{align}
where $P_{PAM, 0}$ is a baseline power consumption that depends on the number of antennas, ganging elements, and the maximum EIRP per antenna; $\eta_{PAM}$ captures the efficiency of the amplifier to convert DC power into signal power gain; and $n_{ant}$ is the number of radio chains (with one amplifier per radio chain). Additionally,  $ P_{PAM, c}(r)$ is a factor that depends on the amount of subcarriers to be amplified, which in turn depends on the number of RBs $r$ carried therein. 

\begin{figure}[t!] 
\centering
\includegraphics[width=0.97\columnwidth]{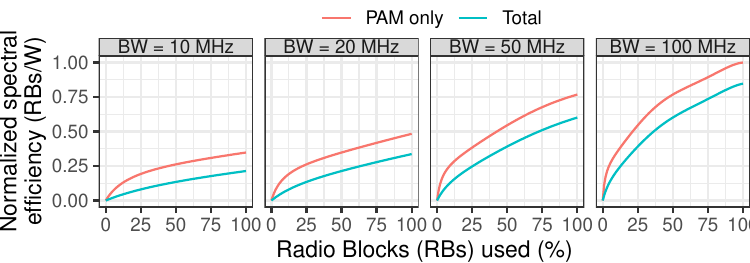}
  \vspace{-4mm}
\caption{Empirical results of spectral efficiency (RBs transmitted per watt of power consumed) of an O-RAN 7.2-x compliant 32TRX massive MIMO RU.}
  \vspace{-4mm}
\label{fig:motivation:ru_pwr}
\end{figure}
Fig.~\ref{fig:motivation:ru_pwr} presents experimental measurements on the RU configured with different bandwidths. The figure depicts the normalized spectral efficiency, measured as the number of radio blocks (RBs) transmitted per watt of power consumed in one symbol period. We present results for the complete RU ($P$ in eq.~\ref{eq:ru_power}) and for the PAM alone ($P_{PAM}$ in eq.~\ref{eq:ru_power}). As mentioned previously, the consumption of the remaining components is independent of the load. 

As expected, overall RU consumption is dominated by the PAM. Moreover, while the PAM's spectral efficiency plateaus as the utilized spectrum resources (RBs) increase, it is maximized when more bandwidth is employed. This observation motivates the principle behind the solution we propose in this paper: \emph{consolidate data into as many radio resources as possible within as few symbol periods as possible.}

\subsection{5G energy-saving features}\label{sec:background:asm}

3GPP has introduced a series of energy-saving features. 
Early on, 3GPP introduced a \emph{lean carrier design}~\cite{dahlman20205g}, which enables grouping together the BS's broadcast signals into periodic Synchronization Signal Blocks (SSB). The SSB periodicity can be adjusted by the operator in the range of 5 to 160~ms, to enable longer intervals for BS sleeping. Moreover, inherited from 4G, Discontinuous Transmission (DTX)~\cite{3GPP_TS_38_321} allows the transmitter to temporarily pause when not sending data, conserving power during the aforementioned small timescale idle periods.

\begin{table}[t!]
\centering
\caption{\footnotesize Advanced Sleep Modes (ASM) considered in this paper.}
\vspace{-2mm}
\label{table:asm}
\begin{tabular}{l|l|c|c} 
                    & \textbf{Off RU HW}                    & \textbf{Normalized}       &    \textbf{Switching}                               \\
\textbf{RU mode}    &  (see \S\ref{sec:analysis:power}) & \textbf{power consumption} &  \textbf{delay ($\mu$s)}  \\ \hline 
Idle & - & 1 & 0 \\ 
ASM 1 & PAM & 0.675 & 37 \\ 
ASM 2 & PAM, BB & 0.55 & 500 \\ 
ASM 3 & PAM, BB, RF & 0.23 & 5000 \\
\end{tabular}
\vspace{-4mm}
\end{table}

Advanced Sleep Modes (ASMs) enable progressively deeper sleep levels by deactivating more RU components, with the trade-off of increased wake-up/sleep switching delays. Based on our analysis in \S\ref{sec:analysis:power} and the related literature~\cite{asm_multi_agent}, we establish the ASMs described in Table~\ref{table:asm}. Our RU already implements ASM 1 by deactivating the PAM, saving over 30\% energy consumption compared to the active (yet idle) state. It requires one OFDM symbol period ($\sim\!\!37~\mu$s with numerology $\mu=1$) for switching, aligning with existing DTX literature reporting PAM deactivation times of $30-65~\mu$s~\cite{tombaz2014energy, chatzipapas2011minimization}. 

We simulate the remaining modes (ASM 2 and ASM 3) as follows. As shown in Table~\ref{table:asm}, we assume that ASM 2 further deactivates the digital processor cores and other  PHY layer components (BB in eq.~\ref{eq:ru_power}). Consequently, we adopt a $0.5$~ms switching delay, consistent with the literature~\cite{salem2019traffic, asm_multi_agent, asm_q_learning, asm_tradeoff_energy_delay} and reported to C-state transitions in Intel processors~\cite{smejkal2023sleep}. Moreover, we assume that ASM 3 further deactivates RU components such as ADC/DAC and frequency converters (RF in eq.~\ref{eq:ru_power}), with a $5$~ms switching delay based on the literature~\cite{salem2019traffic, asm_multi_agent, asm_q_learning, asm_tradeoff_energy_delay}.

\begin{figure}[t!] 
\centering
\includegraphics[width=0.9\columnwidth]{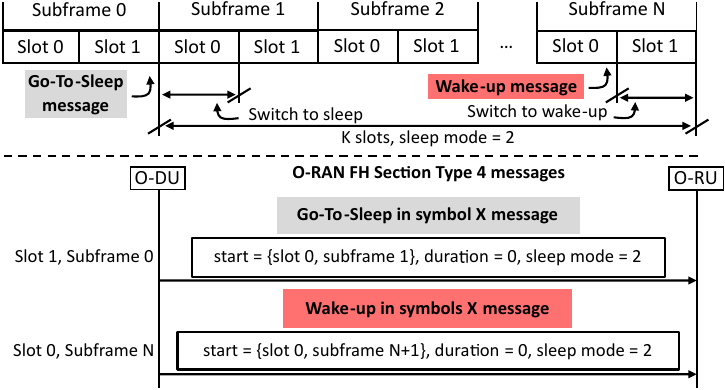}
  \vspace{-3mm}
\caption{Using O-RAN Section Type 4 messages to implement ``Go-to-sleep'' and ``wake-up'' commands. Example for ASM 2 and numerology $\mu=1$.}
  \vspace{-6mm}
\label{fig:analysis:o_ran_example}
\end{figure}

To asynchronously wake up or put the RU to sleep, \name relies on O-RAN's  open fronthaul (FH) specification~\cite{oran-asm}.  O-RAN uses ``Section Type 4'' messages within the FH’s C-Plane to configure specific ASMs in an RU. These C-Plane messages convey slot-level configurations, including ASM settings, and apply to all endpoints associated with a carrier, antenna array, or the entire RU. Fig.~\ref{fig:analysis:o_ran_example} shows an example for numerology $\mu=1$. To activate ASM 2, the DU sends the RU a ``Go-to-sleep'' command with a ``Section Type 4'' message with the following parameters: ($i$) a symbol mask indicating the symbol period within the next slot when the RU should enter sleep mode, ($ii$) ``start'' in the next immediate slot, ($iii$) duration set to 0 (undefined), and ($iv$) the ASM identifier. Conversely, to wake up the RU, the DU sends a ``wake-up'' message by \emph{unmasking} the previously sent ``Go-to-sleep'' message.

%% file: mechanism_design.tex
\section{\name Design}\label{sec:design}

The analysis in \S\ref{sec:exp_analysis} highlights the diverse range of RU sleeping opportunities and the inherent trade-off between cell idle times and energy savings. 

An additional observation is that 5G network slices have varying user delay requirements. For instance, while Ultra-Reliable Low-Latency Communication (URLLC) slices demand the lowest latency for critical applications, targeting delays of less than 1~ms, enhanced Mobile Broadband (eMBB) slices tolerate moderate delays for smooth streaming, aiming for 1-10~ms, or Massive Machine-Type Communication (mMTC) slices, designed for massive IoT deployments, prioritize energy efficiency over strict latency requirements, often accepting delays of 10~ms and above. This observation presents an additional opportunity for energy savings: to trade off data delay for deeper and longer sleep intervals, \emph{provided that slice QoS requirements are met}. 

However, attaining such guarantees requires a radio scheduler at the MAC layer that is both RU-aware (to exploit sleeping opportunities) and QoS-driven. As we discuss in \S\ref{sec:related}, a large body of research studies schedulers with QoS guarantees (e.g.,~\cite{ji2023downlink}), but ignore small-timescale RU energy-saving opportunities. Some recent work optimizes Advanced Sleep Modes (ASM) for this purpose, but they do not provide QoS guarantees (e.g.,~\cite{asm_multi_agent}). More importantly, very few of these works are actually practical for real-world base stations due to the tight computing constraints in the DU~\cite{gpf}. 

\begin{figure}[t]
    \centering
    \includegraphics[width=\columnwidth]{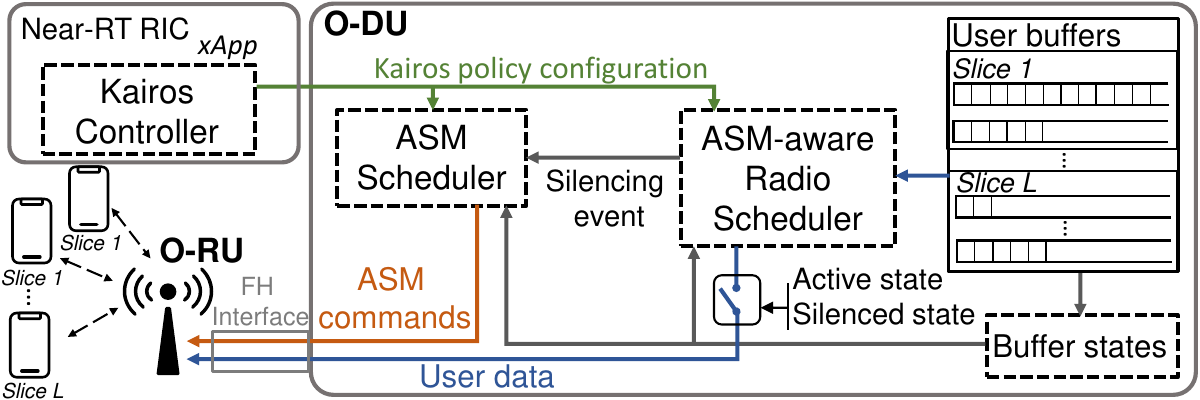}
    \vspace{-6mm}
    \caption{{\name system architecture.}}
    \vspace{-6mm}
    \label{fig:scheme}
\end{figure}

To address these challenges, we propose a novel approach called \name, which, unlike prior work, leverages the O-RAN-enabled interplay between real-time MAC-layer procedures (\S\ref{sec:design:mac}) and near-real-time radio-policing xApps (\S\ref{sec:design:xapp}). This two-tier O-RAN-compliant framework is easy to implement and effectively exploits ASMs to save energy while guaranteeing hard QoS requirements. As depicted in Fig.~\ref{fig:scheme}, \name comprises three components: the ASM-aware radio scheduler and the ASM scheduler, operating in real-time within the DU, and the \name controller, operating in near-real time within the Near-RT RIC, which are presented next.

\subsection{Real-time DU operation: \name policy}\label{sec:design:mac}
 
\subsubsection{ASM-Aware Radio Scheduler}

Our goal is to design a minimally intrusive radio scheduling policy that preserves the operator's intended scheduling logic and is lightweight enough to avoid violating MAC layer computing constraints, while aiding in extending RU sleep cycles within predefined QoS requirements. To this end, the \name policy establishes two states that a legacy MAC-layer scheduler must abide by:
 \begin{itemize}[noitemsep,topsep=0pt,parsep=0pt,partopsep=0pt,leftmargin=8pt]
    \item \textbf{Active}: The radio scheduler follows its own logic to allocate radio resources to users. However, the aggregated user data \emph{must} be consolidated into as few OFDM symbols (time resources) as possible, maximizing the amount of RBs allocated per symbol period (spectrum resources). When the data buffer is empty, a silencing event is triggered.
    
    \item \textbf{Silenced:} The radio scheduler is not allowed to allocate radio resources, and downlink data is buffered. It also tracks the waiting time or ``age'' of the oldest user data burst in each slice, computed by the buffer states module (Fig.~\ref{fig:scheme}). If any burst's age exceeds the \name policy configuration $d$, then an \textit{activating} event is triggered. More formally, let $\mathcal{L}$ be the set of network slices. We define $\mathcal{B}^{(l)}_t$ as the set of user data bursts from slice $l \in \mathcal{L}$ waiting in the downlink RLC buffers at time $t$. Let $d_t$ be the \name policy configuration at time $t$. Thus, an \emph{activating} event is triggered whenever $age(b) > d_t, \forall b \in \mathcal{B}^{(l)}_t, \forall l \in \mathcal{L}$. 
\end{itemize}

\begin{figure}[t!] 
\centering
\includegraphics[width=0.76\columnwidth]{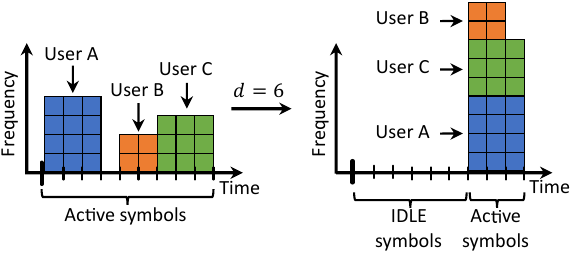}
\vspace{-4mm}
\caption{{Rationale behind \name policy.}}
\vspace{-3mm}
\label{fig:policy_motivation}
\end{figure}

Note that downlink transmissions shall resume in the symbol period immediately following an \emph{activating} event, at which time all BS components should be awake.  As shown in  Fig.~\ref{fig:policy_motivation} for a toy example with a policy equal to $d=6$ OFDM symbols, delaying small traffic bursts helps consolidating radio blocks into fewer OFDM symbols, which improves energy efficiency (see Fig.~\ref{fig:motivation:ru_pwr}).

\subsubsection{ASM Scheduler}

\begin{figure}[t]
\vspace{-1mm}
    \centering
    \includegraphics[width=0.95\columnwidth]{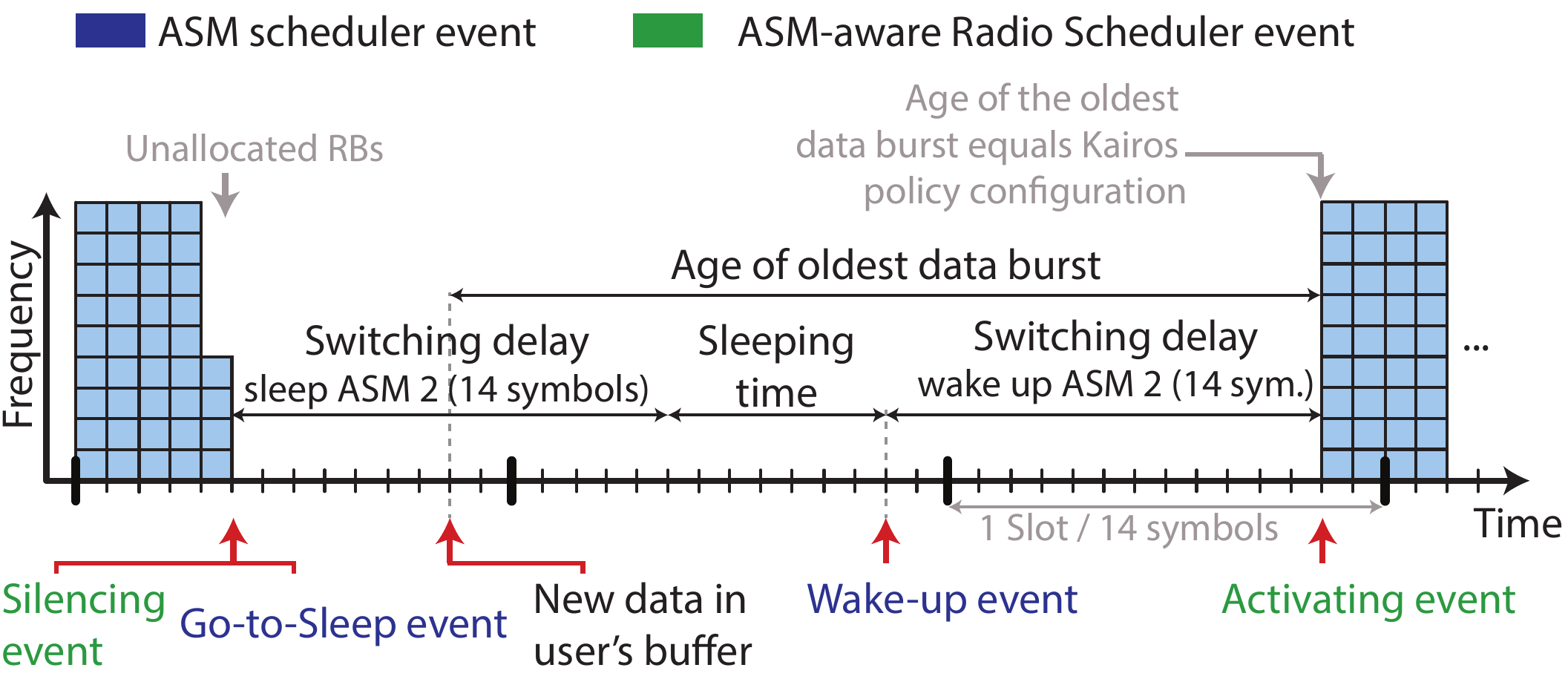}
    \vspace{-4mm}
    \caption{Example of a sleeping period using \name policy $d=28$ symbols. }
    \vspace{-5mm}
    \label{fig:example_operation}
\end{figure}

The \textit{ASM scheduler} generates ASM control commands based on some logic. Prior work has proposed somewhat complex ASM control algorithms (e.g.,~\cite{asm_multi_agent}). However, these approaches often overlook practical issues concerning computing and real-time constraints. Our goal is to devise a simple solution, suitable for real-time BS operation, that leverages the aforementioned policy to fully exploit the potential of ASMs for downlink transmissions.

When the ASM-aware radio scheduler triggers a silencing event, it notifies the ASM scheduler (Fig.~\ref{fig:scheme}). This component then selects an ASM based solely on the \name policy configuration. We propose a simple method: to select the deepest ASM whose total switching delay (see Table~\ref{table:asm}) is lower than the \name policy configuration. In other words, the duration of the sleep mode should \emph{fit} within the silenced interval provided by the \name policy. For instance, in a scenario with a \name policy configuration of $d=3$ ms, the selected sleep mode would be ASM 2. To configure the selected ASM in the RU, we use the O-RAN FH interface, as explained in \S\ref{sec:background:asm}.  Note that the RU must be awake during the symbol period immediately following an activation event triggered by the radio scheduler. Therefore, the ASM scheduler must notify the RU to wake up in advance, with the lead time depending on the specific sleep mode.

Fig.~\ref{fig:example_operation} illustrates the operation of the \name policy, where the x-axis represents time resources (symbol periods and slots), and the y-axis depicts radio resources (RBs). Unused RBs in the fifth symbol trigger a ``Silencing'' event in the radio scheduler, leading to a ``Go-to-sleep'' command in the ASM scheduler. Since the \name policy is configured with $d=1$~ms (2 slots), ASM 2 is selected because it is the deepest mode with a total switching delay less than $d$ (see Table~\ref{table:asm}). Seven symbol periods later, new data arrives while the RU is asleep. After 2 slots, the age of this data burst shall meet the policy configuration, which should cause the radio scheduler to trigger an ``Activating'' event. Therefore, a ``wake-up'' message is scheduled one slot before the ``Activating'' event, aligning with ASM 2's one-slot switching delay.

\begin{figure}[t!] 
\centering
\includegraphics[width=0.8\columnwidth]{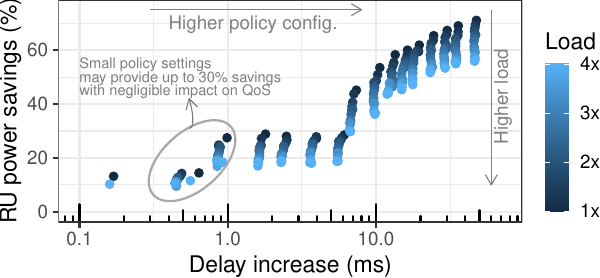}
    \vspace{-3mm}
    \caption{Analysis of \name radio scheduling policy.}
    \vspace{-6mm}
\label{fig:policy_analysis}
\end{figure}

\subsubsection{Analysis}

In this way, the \name policy enables selecting a Pareto-optimal operating point tailored to the requirements of a given network slice (policy optimization is addressed later). Fig.~\ref{fig:policy_analysis} illustrates the trade-off between RU power savings (y-axis) and extra delay incurred (x-axis) for various \name policy configurations and load levels. We adopt a trace-driven approach, using the data from real cells that we presented in \S\ref{sec:exp_analysis}, to emulate load on the same O-RAN 7.2-x compliant 32TRX massive MIMO RU employed in \S\ref{sec:analysis:power}, configured to use 50~MHz of bandwidth and numerology $\mu=1$ (see \S\ref{sec:background:primer}). We define ``1x'' load as the load patterns observed in our measurements. To ensure our analysis remains relevant for potential future scenarios with higher cell loads, we generate these higher loads by virtually compressing time in our traces by a factor indicated in the color legend (``2x'', ``3x'', and so on). This method allows us to analyze diverse load levels while preserving the bursty nature of real-world traffic patterns.

The results offer two key insights. First, small policy settings that induce minimal delay increases (0.25-1 ms) enable RU power savings of 30\% (under today's typical loads, i.e,. ``1x'') to 10\% (under loads four times higher than the median cell load today, i.e,. ``4x''). Second, if QoS requirements were more relaxed, permitting 10-40 ms of additional user delay, the \name policy could achieve even greater savings, ranging from 70\% (under ``1x'' loads) to 35\% (under higher loads).

\vspace{-1mm}
\subsection{Near-RT RIC xApp: \name controller}\label{sec:design:xapp}

The \textit{\name Controller} is essentially an O-RAN xApp running at the Near-RT RIC that dynamically computes the policy configuration (threshold delay) over time. As shown in \cite{fiore_spatiotemporal}, different services (hence, slices) show different temporal load patterns even if they belong to the same category. Hence, it is of paramount importance to dynamically adjust the policy configuration, considering the behavior of all the slices, in order to harness the maximum possible energy savings while guaranteeing QoS requirements.
This is a challenging task as the user's experienced delay does not only depend on the \name{} policy but also on the traffic profile that the BS has to handle (i.e., traffic load and type of other users, other network slices, etc). We address this in detail in \S\ref{sec:algorithm}.

%% file: policy_opt.tex
\section{\name controller}\label{sec:algorithm}

Next, we first formally describe the policy control problem (\S\ref{sec:algorithm:problem}) and then propose a solution (\S\ref{sec:algorithm:solution}).

\subsection{Problem formulation}\label{sec:algorithm:problem}

As explained above, the \name controller is responsible for adjusting the \name policy configuration used by the real-time modules (ASM-aware radio scheduler and ASM scheduler) based on changes in the traffic demands and different QoS requirements across slices.
Following the O-RAN architecture, the \name controller operates in the Near-RT RIC, i.e., with a time granularity of $\sim\!\!100$~ms. Hence, we define the time steps denoted by $t = \{0, 1,\ldots\}$ based on this time granularity.

We let $d_t \in \mathcal{D}$ denote the \name policy configuration at time step $t$. We do not allow different policies for different slices for two reasons: ($i$) learning a separate policy per slice from data would exponentially increase the time and data required for convergence to a good model; and ($ii$) although different slices may have different user delay requirements, the physical RU is shared. Thus, if the policy for one slice triggers the scheduler into an ``Active'' state, there is no energy-saving benefit from deferring transmissions on other slices, as the RU would already be active.

Furthermore, we consider a BS with a set $\mathcal{L}$ of $L_{\text{max}}$ network slices. We do not make any assumption on the distribution of users across slices.
Given that the number of active slices in the network can change over time, we denote by $\mathcal{L}_t \subseteq \mathcal{L}$ the set of active slices at time $t$, whose cardinality is $|\mathcal{L}_t| = L_t$.
The target requirement of slice $l \in \mathcal{L}$ is denoted by $\Delta^{(l)}$.

We now let $\mathcal{I}_t^{(l)}$ describe the set of data bursts generated by the users in slice $l \in \mathcal{L}$ during the time step $t$. Each data burst $I \in \mathcal{I}_t^{(l)}$ is defined by its size (in bits) and the generation time. We characterize the user data bursts using the distribution of their inter-arrival time (IAT) and size. Specifically, for a given set $\mathcal{T} \in [0,1]^N$ of $N$ quantiles, we define $\zeta_t^{(l)} = \Phi^{\text{IAT}}(\mathcal{I}^{(l)}_t, \mathcal{T})$ as the vector of quantile values of the IAT distribution. Similarly, $\xi_t^{(l)} = \Phi^{\text{size}}(\mathcal{I}^{(l)}_t, \mathcal{T})$ denotes the vector of quantile values of the burst size distribution. Then, we define the context of slice $l$ at time $t$ as $s_t^{(l)}:= (\zeta_t^{(l)}, \xi_t^{(l)})$. Now, we can define the joint context of the BS by $s_t = \{s_t^{(l)}\}_{l \in \mathcal{L}_t} \in \mathcal{S}$.

We now define the energy consumption of the system as $E(s_t, d_t)$. Since all slices operate over the same RU, the consumed energy is a global metric. We define the QoS metric per slice as $\delta^l(s_t, d_t)$. For example, this metric can be related to the delay experienced by users. Finally, we define the function $\pi : \mathcal{S} \mapsto \mathcal{D}$ that maps joint contexts to \name policy configurations. Consequently, our objective is to solve the following contextual bandit problem:

\begin{problem}\label{eq:main_prob}
\begin{align}
  & \operatorname*{argmin}_{\pi} \sum_{t=1}^T E(s_t, \pi(s_t))  \nonumber& \\
  & \textup{s.t.} \quad \delta^{(l)}(s_t, \pi(s_t)) < \Delta^{(l)}, & \forall l &\in \mathcal{L}_t, t = 1, \ldots, T.  \nonumber 
  \end{align}
\end{problem}

Problem~\ref{eq:main_prob} aims to minimize the average energy consumption of the RU while satisfying the  \textit{hard}  QoS constraints of every slice at each time step. 

We would like to emphasize that the problem in \eqref{eq:main_prob} is challenging due to several factors. 
Both the energy $E(\cdot)$ and the QoS per slice $\delta^{(l)}(\cdot)$ are complex functions that depend not only on $s_t$ and $\pi(\cdot)$, but also on other system parameters (e.g., bandwidth). Intuitively, higher values of $d_t$ can reduce power consumption by enabling longer silent periods. However, the value of the QoS metric $\delta^{(l)}(\cdot)$ is complex to devise in advance as it depends on the traffic and buffer states of all other slices. Moreover, these metrics can exhibit randomness due to unknown and uncontrollable environmental factors, which must be considered to satisfy the hard constraints. Furthermore, the number of active slices can vary over time, increasing the problem's complexity and precluding the use of standard machine learning approaches (with fixed input size and number of constraints). We next overcome these challenges with a novel learning framework.

\subsection{A novel ML framework for ASM control}\label{sec:algorithm:solution}

Our solution is designed to overcome the next challenges:
\begin{itemize}
\item \textit{Variable input size:} The size of the context $s_t$ may change over time due to the varying number of active slices. Hence, we propose a dimensionality-invariant encoder that projects the context into a fixed dimensional space. 
\item \textit{Hard constraints with uncertain QoS:} Problem~\eqref{eq:main_prob} imposes hard constraints that are difficult to satisfy, as QoS measurements in the network $\delta^{(l)}(\cdot)$ can exhibit randomness due to unknown and uncontrollable factors. To address this, we use distributional critics to capture not only the mean value of the function $\delta^{(l)}(\cdot)$ but also its distribution. Thus, by examining the tail of the distribution, we can ensure constraint satisfaction with some pre-determined probability.
\item \textit{Variable number of constraints:} The problem involves a potentially large and variable number of constraints. To address this, we propose an ML architecture with one actor (to select $d_t$) and multiple distributional critics (one per constraint). This model allows us to compute a joint cost function and backpropagate the loss through all active critics (based on the number of active slices), with the goal of learning the optimal actor function.
\end{itemize}

In the following, we detail each component of our solution.

\subsubsection{Dimensionality-invariant encoder}
As mentioned previously, the number of active slices $L_t$ can be potentially large and vary over time. A naive approach to inputting $s_t$ into the machine learning framework would be to concatenate $s_t^{(l)}$ $\forall l \in \mathcal{L}_t$ and then use a single neural network (NN) with the appropriate input size. However, this requires a separate NN for each possible combination of active slices, totalling  $\sum_i \binom{L_{\text{max}}}{i}$ encoders, as the input is not permutation invariant. In other words, the traffic of each slice $l$ is associated with a specific target QoS $\Delta^{(l)}$.

Instead, we propose a much simpler, scalable, and effective solution inspired by relational networks~\cite{relational_networks} and graph neural networks~\cite{gat}. We project the context of each slice into a higher-dimensional space and then combine the projections of all slices using an aggregator function. While summation is a common choice for aggregation in the literature, it is permutation invariant. To differentiate between contexts corresponding to different slices, we incorporate the slice identifier into the projection function. Thus, the encoded context representation with a fixed number of dimensions is given by:
\begin{equation}\label{eq:encoder}
    \tilde{s}_t = \sum_{l \in \mathcal{L}_t} g (s_t^{(l)}, l | \phi),
\end{equation}
where $g (\cdot | \phi)$ is the projection function with parameters $\phi$. 

\subsubsection{Distributional critics}
The main idea behind our distributional critics is to capture the distribution (i.e., the value of a set of quantiles) of the target function. This enables us to precisely identify the tail of the distribution and thus satisfy constraints with certain probability.

For a given random variable $X$, the value of its quantile function is defined as $q_\tau = F^{-1}_X(\tau)$, where $\tau \in [0, 1]$ is the quantile and $F_X(x)$ is the cumulative distribution function. We employ the quantile regression loss,  an asymmetric convex function that penalizes underestimation error with weight $1-\tau$ and overestimation error with weight $\tau$: 
 \begin{align}\label{eq:qr_loss}
    \mathcal{J}^\tau (\hat{q}_\tau) &:= \mathbb{E}_{x\sim X} \left[ \rho_\tau (x - \hat{q}_\tau) \right], \text{where} \\
    \rho_\tau (u) &:= u\cdot (\tau - \delta_{\{u<0\}}) \;\;\;\; \forall u \in \mathbb{R},
\end{align}
where $\hat{q}_\tau$ is the estimated value of the quantile function, and $\delta_{\{z\}}$ is an indicator function that takes the value $1$ when $z$ holds and $0$ otherwise.

For each critic, we approximate $N$ quantile values $q_{\tau_1}, \ldots, q_{\tau_N}$ by training the critic using gradient decent to minimize the following objective:
\begin{equation}\label{eq:loss_all}
    \sum_{i=1}^N \mathcal{J}^{\tau_i} (\hat{q}_{\tau_i}).
\end{equation}

Since the loss function defined above is not smooth at $u=0$, the performance of function approximators (e.g., NNs) may not be optimal. To address this, we use the \textit{quantile Huber loss}~\cite{huber}. This loss function, instead, presents a squared shape in the interval $[-\kappa, \kappa]$, and reverts to the standard quantile loss outside this interval:
\begin{equation}
 J_\kappa (u) :=
  \begin{cases}
    \frac{1}{2} u^2       & \quad \text{if } |u| \leq \kappa \\
    \kappa (|u| - \frac{1}{2} \kappa )  & \quad \text{otherwise.} 
  \end{cases}
\end{equation}

The asymmetric version of the Huber loss is given by:
\begin{equation}\label{eq:huber}
    \rho^\kappa_\tau (u) := |\tau - \delta_{\{u<0\}}| \frac{J_\kappa (u)}{\kappa}.
\end{equation}
By substituting $\rho^\kappa_\tau (u)$ into eq~\eqref{eq:qr_loss}, we obtain the quantile Huber loss. Note that as $\kappa \to 0$, the quantile Huber loss reverts to the quantile regression loss.

\subsubsection{Framework architecture and learning procedure}

To handle a large and variable number of constraints in our problem, we use an architecture with one actor and $L_{\text{max}}+1$ distributional critics \cite{ayala2024risk}. 

The actor function $\pi(\tilde{s} | \eta)$, parameterized by $\eta$, can take continuous values as the \name policy is continuous. This type of actor is commonly referred to as a deterministic actor~\cite{dpg}.
The distributional critics are denoted by $C^{(l)}(\tilde{s}, d | \theta^{(l)})$, where $\theta^{(l)}$ represents the parameters of critic $l$. We use the index $l=0$ for the critic that approximates the objective function (energy consumption in \eqref{eq:main_prob}) and $l=1, \ldots, L_{\text{max}}$ for the critics that approximate the constraint of slice $l$, i.e., $\delta^{(l)}(\cdot)$.

We define an aggregate cost signal to capture the information about all constraints provided by the critics:
\begin{align}\label{eq:ragg}
    & C^{agg}_t(\tilde{s}_t,d_t, \mid \theta)  := \\
    & \bar{C}^{(0)}(\tilde{s}_t,d_t \! \mid \! \theta^{(0)}) \! + \! \sum_{l \in \mathcal{L}_t} \lambda \! \max\left( \! \gamma^{\alpha} \! (C^{(l)}(\tilde{s}_t,d_t \! \mid \! \theta^{(l)})) \! - \! \Delta^{(l)} , \! 0 \! \right) \! , \nonumber
\end{align}
where $\gamma^{\alpha}(X)$ is the the quantile function value of distribution $X$ at quantile $\alpha$, $\lambda$ is the penalty weight for the constraints, $\theta = \{\theta^{(0)}, \ldots, \theta^{(L_{\text{max}})}\}$ is the joint set of parameters of the $L_{\text{max}}+1$ critics, and $\bar{C}^{(l)}(\cdot)$ is the mean of the distribution of critic $l$.

In detail, the first term in eq.~\eqref{eq:ragg} represents the mean of the objective function (energy) that we aim to minimize, while the second term aggregates the penalty incurred by the constraint violations. When a constraint is satisfied, the value inside the $\max()$ function is negative and the penalty is zero.
Importantly, the use of the quantile function in the aggregated cost function ensures that the tail of the distribution of the constraints (as $\alpha \rightarrow 1$) meets the constraints, boosting the robustness of our solution in terms of constraint satisfaction. 
While this formulation focuses on QoS metrics that should be below a maximum value (e.g., delay), eq.~\eqref{eq:ragg} can be easily adapted for QoS metrics that have a minimum target (e.g., throughput) by selecting values of $\alpha$ close to zero and reversing the sign inside the maximum operator.

To train the actor, we first define its objective function as
\begin{equation}\label{eq:R}
R(\pi) = \mathbb{E}_{\tilde{s} \sim \beta} [C^{agg}(\tilde{s}, \pi(\tilde{s} \mid \eta) \mid \theta)]
\end{equation}
where $\beta(\tilde{s})$ is the stationary distribution of the projected contexts. 
Note that in a contextual bandit problem, the distribution of the contexts is not conditioned by the actor function.

Then, we derive the actor update by applying the chain rule to the performance objective defined in eq.~\eqref{eq:R} with respect to the actor parameters~\cite{dpg}:
\begin{equation}
    \nabla_{\eta} R(\pi) \approx \mathbb{E}_{\tilde{s} \sim \beta} \left[ \nabla_{a} C^{agg}(\tilde{s}, d \mid \theta)\mid_{d=\pi (\tilde{s} \mid \eta)} \nabla_{\eta} \pi (\tilde{s} \mid \eta) \right].
\end{equation}

\begin{figure}[t]
\vspace{-2mm}
    \centering
    \includegraphics[width=\columnwidth]{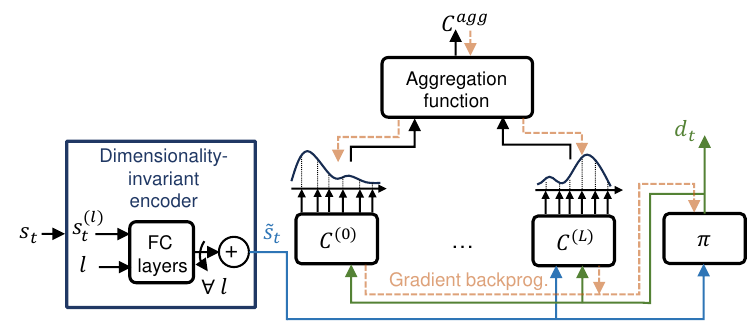}
    \vspace{-6mm}
    \caption{\name controller comprising a dimensionality-invariant encoder, a deterministic actor, $L+1$ critics, and an aggregation function (eq.~\eqref{eq:ragg}).}
    \vspace{-2mm}
    \label{fig:ml_scheme}
\end{figure}

For a practical implementation of the algorithm, we also consider a replay buffer $\mathcal{D}$ to store samples of experience from each time step, as in \cite{dqn}.
Gradients are then computed using mini-batches of $B$ samples randomly drawn from the replay buffer.
Empirically, we observed that the distributional critics achieve better performance when approximating multiple quantiles rather than just $\alpha$. Therefore, we define $T$ as the set of quantiles approximated by the critics, where $\alpha \in T$. At each time step $t$, we add noise $\mathcal{N}_t$ to the output of the actor to promote exploration during training \cite{ddpg}. Fig.~\ref{fig:ml_scheme} illustrates all the components of the \name controller and Algorithm~\ref{alg:algorithm} presents its training procedure.

In this way, \name is purposely designed to maximize energy efficiency, consolidating radio resources in as few OFDM symbols as possible, within a deterministic delay bound. This is in contrast with the literature's ASM-only strategies, which are unable to provide hard QoS guarantees. 

\setlength{\textfloatsep}{5pt}
\begin{algorithm}[t!]
    \caption{\name controller training}
    \footnotesize
    \label{alg:algorithm}
    \textbf{Input}: $\alpha$, $\kappa$, $B$, $T$\\
    \textbf{Initialize}: Parameters $\eta$, $\theta$; reply buffer $\mathcal{D} = \emptyset$;\\ Exploration noise $\mathcal{N}$; 
    \begin{algorithmic}[1]
        \FOR{$t = 1, \ldots, T$}
            \STATE Observe the context of all active slices $s_t$
            \STATE Compute $\tilde{s}_t$ using eq.~\eqref{eq:encoder}
            \STATE Compute the \name policy configuration $d_t = \pi(\tilde{s}_t, \mid \eta) + \mathcal{N}_t$
            \STATE Apply the \name policy configuration during the time step $t$
            \STATE At the end to step $t$, observe the energy $E_t$
            \STATE Observe the QoS per slice $\delta_t = \{ \delta^{(l)}_t \}_{l \in \mathcal{L}_t}$
            \STATE Store in $\mathcal{D}$ the experience sample $\langle \tilde{s}_t, d_t, E_t, \delta_t \rangle$
            \STATE Sample a random minibatch of $B$ samples $\langle \tilde{s}_i, d_i, E_i, \delta_i \rangle$
            
            \STATE Update the critic $0$ (energy) by minimizing the loss \\ $\frac{1}{B} \sum_i \sum_{\tau \in T} \rho^\kappa_{\tau} (E_i - \gamma^{\tau}(C^{(0)}(\tilde{s}_i,d_i | \theta^{(0)})))$

            \FOR{ $l \in \mathcal{L}_t$}
                \STATE Update the critic $l$ by minimizing the loss \\ $\frac{1}{B} \sum_i \sum_{\tau \in T} \rho^\kappa_{\tau} (\delta_i^{(l)} - \gamma^{\tau}(C^{(l)}(\tilde{s}_i,d_i | \theta^{(l)})))$
            \ENDFOR
            
            \STATE Update the actor with the sampled policy gradient \\ $\frac{1}{B} \! \sum_i \! \! \nabla_{d} \; C^{agg}  (\tilde{s}_i,  d,  | \theta)|_{d=\pi (\tilde{s}_i | \eta)}  \nabla_{\eta} \pi (\tilde{s}_i | \eta  )$ 
        \ENDFOR
    \end{algorithmic}
\end{algorithm}

%% file: perf_eval.tex
\section{Performance Evaluation}\label{sec:perf}

To evaluate \name, we utilize the commercial RU and the emulator of real-world traces introduced in \S\ref{sec:exp_analysis}. Using this testbed, we implement \name{} as described in \S\ref{sec:design} and \S\ref{sec:algorithm}, and evaluate its performance in terms of convergence (\S\ref{sec:perf:convergence}), in comparison with alternative radio and ASM scheduling solutions (\S\ref{sec:perf:comparison}), and its performance in dynamic environments with time-varying network slices (\S\ref{sec:perf:dynamic}).

Moreover, \name is configured with $\alpha=0.995$, a reply buffer $\mathcal{D}$ of $10^4$ samples, and batch size $B=128$. The training exploration noise $\mathcal{N}$ follows an Ornstein-Uhlenbeck process with parameters $\theta_{\text{noise}} = 0.15$ and $\sigma_{\text{noise}} = 0.15$ \cite{ddpg}. Finally, we consider a decision time step of $200$~ms.

\subsection{Convergence}\label{sec:perf:convergence}

\begin{figure}[t!] 
\centering
\includegraphics[width=\columnwidth]{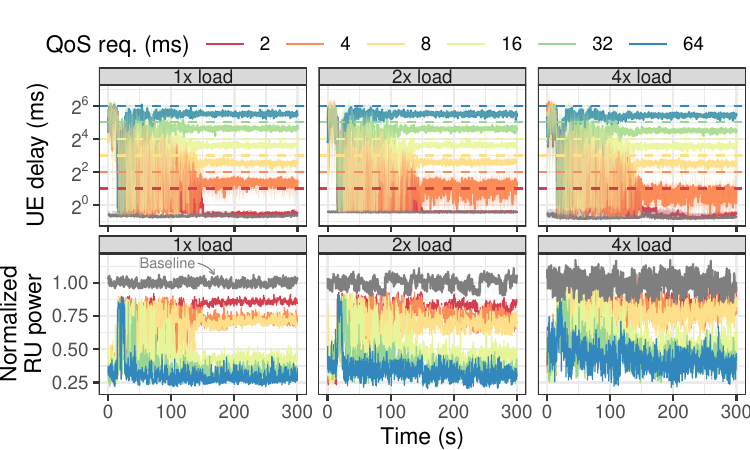}
\vspace{-6mm}
\caption{\name convergence. The dashed lines represent the QoS requirement. The grey lines represent the performance of an ASM-unaware baseline.}
\label{fig:perf:convergence}
\end{figure}

To study the convergence of \name, we test the traces from \S\ref{sec:exp_analysis} under various load levels, starting without prior training. We repeat the experiment for different maximum UE delay QoS requirements. Fig.~\ref{fig:perf:convergence} depicts the evolution over time of the mean user delay (top) and normalized RU power consumption\footnote{Power consumption figures are normalized to respect the Non-Disclosure Agreement with equipment vendor.} (bottom) for each scenario. 

\name consistently meets the QoS target after approximately 150 seconds of training (750 time steps). To this end, \name induces higher RU power consumption for stricter QoS requirements. For example, under a ``4x'' load, the mean power consumption is 0.39 when $\Delta=64$ ms, but increases to 0.85 (over twofold) when $\Delta=2$ ms. This is expected, as stricter QoS targets require less spectrally efficient allocations  as those discussed in \S\ref{sec:analysis:power}. However, our approach achieves  power savings that range between 72\% (``1x'' load and $\Delta=64$~ms) and 15\% (``4x'' load and $\Delta=2$~ms) compared to an ASM-unaware baseline  (depicted in grey).

We observe that user delay, our measure of QoS performance, more closely approaches its target as the target becomes less strict. This is because the bursty traffic behavior of our traces smoothens out with higher \name policy settings, as more data bursts are buffered. This behavior increases traffic predictability and allows \name to operate more aggressively, saving energy while still meeting the QoS target. Conversely, tighter QoS targets limit the radio scheduler's ability to buffer and smooth out data bursts, requiring a more conservative operation to ensure QoS guarantees are met.

\subsection{Comparison}\label{sec:perf:comparison}

We next compare \name with different benchmarks.

\subsubsection{ASM scheduler} 

\begin{figure}[t!] 
\centering
\includegraphics[width=0.65\columnwidth]{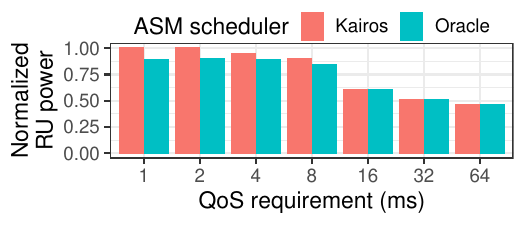}
\vspace{-4mm}
\caption{\name ASM scheduler \emph{vs} an Oracle solution for ``4x'' load.}
\vspace{-2mm}
\label{fig:perf:comparison_asm}
\end{figure}

We first evaluate \name's ASM scheduler against an oracle, both operated by \name controller.
This oracle ASM scheduler is an idealized approach with perfect foresight of future data burst arrivals, allowing it to make the optimal ASM schedule. Fig.~\ref{fig:perf:comparison_asm} depicts the power consumption achieved by both approaches for a ``4x'' load scenario (other load levels yield similar results). Notably, the figure demonstrates that \name, a simple strategy suitable for real-time operation, performs comparably to the oracle when guided by a near-real-time policy. This contrasts sharply with related, more complex solutions~\cite{salem2019traffic, asm_distributed_q_learning, asm_multi_agent}, which are difficult to implement in real systems due to their computational complexity and the tight computing constraints of DUs.

\subsubsection{\name controller}

\begin{figure}[t!] 
\centering
\includegraphics[width=\columnwidth]{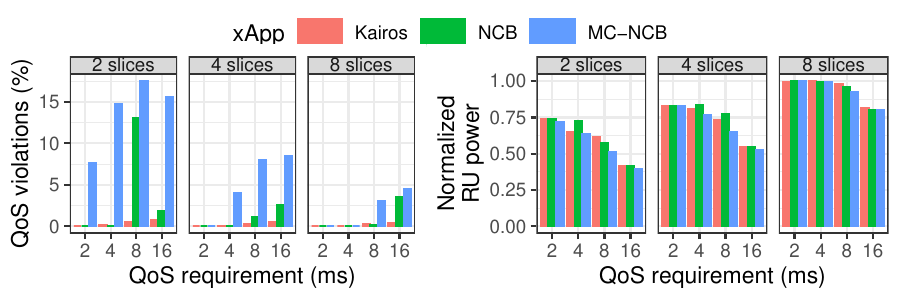}
\vspace{-8mm}
\caption{Comparison between different xApp ML solutions. Load = ``1x''. }
\label{fig:perf:comparison}
\end{figure}

We compare \name controller against two alternative ML architectures that solve our problem:
 \begin{itemize}[noitemsep,topsep=0pt,parsep=0pt,partopsep=0pt,leftmargin=8pt]
 \item \textbf{NCB} (Neural Contextual Bandit) comprises an actor-critic NN architecture \cite{ddpg} modified to fit our problem formulation (see \cite{ayala2024risk}). We define a single utility function where the constraints are Lagrangian-like penalty terms.
 \item \textbf{MC-NCB} (Multi Critic-NCB) extends NCB with one critic per constraint. Each (non-distributional) critic approximates the constraint function using MSE loss. The actor is updated, and the output of all critics is aggregated similarly to eq.~\eqref{eq:ragg}.
\end{itemize}

Fig.~\ref{fig:perf:comparison} depicts the share of data bursts violating their delay target (left plot) and normalized RU power consumption (right plot) for different delay targets (x-axis) and number of slices (2, 4, and 8, respectively), each following the ``1x'' load pattern. As expected, RU power consumption decreases with relaxed QoS requirements and a lower number of slices (i.e., lower aggregated load). However, while RU power consumption is similar across all three approaches, \name significantly outperforms the other two benchmarks in terms of QoS violation rates, achieving nearly negligible violations. Notably, QoS violations are higher with fewer slices (lower aggregated load). This occurs because increasing the load smooths the traffic burstiness, making it more predictable and creating a friendlier environment for the controllers.

\subsection{Dynamic environments}\label{sec:perf:dynamic}

\begin{figure}[t!] 
\centering
\vspace{-2mm}
\includegraphics[width=\columnwidth]{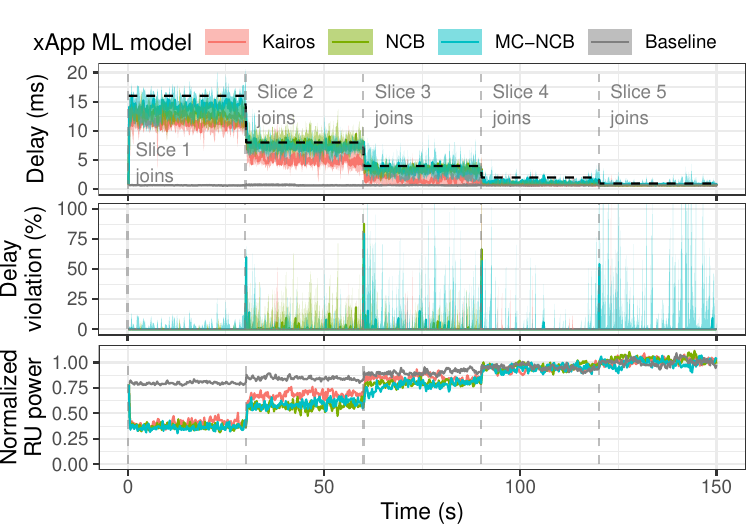}
\vspace{-8mm}
\caption{5 slices with QoS requirements $d=\{16, 8, 4, 2, 1\}$ ms, respectively, joining the system every 30 seconds. The dashed line represents the most stringent QoS requirement.  The grey lines represent the performance of an ASM-unaware baseline. }
\vspace{-1mm}
\label{fig:perf:dynamic}
\end{figure}

We conclude by evaluating \name and the two benchmarks in a more dynamic and heterogeneous setup, emulating a scenario with five network slices (each with distinct QoS requirements of 16, 8, 4, 2, and 1 ms) handling ``1x'' load patterns and joining the system every 30 seconds. All ML models are trained identically. Fig.~\ref{fig:perf:dynamic} shows the burst delay performance (top), excess delay relative to QoS targets (middle), and normalized power consumption (bottom) over time.

All ML models achieve comparable power savings compared to the baseline, with diminishing returns as the number of slices and aggregate load increase, and QoS requirements become stricter (1 ms for slice 5). However, only \name consistently maintains negligible QoS violations across all system states. In contrast, NCB and MC-NCB exhibit significant violations, reaching 19.7\% and 51\% at the 99th percentile, respectively. These violations are particularly pronounced during state transitions as new slices join, whereas \name seamlessly adapts to new slices without incurring additional delays.

%% file: related_work.tex
\section{Related Work}\label{sec:related}

The concept of cell sleeping for energy saving has been extensively explored in the literature~\cite{cell_sleeping_survey}. For instance, the authors in~\cite{cell_sleeping_1} address the joint problem of cell clustering and BS sleeping, while~\cite{cell_sleeping_3} proposes a joint strategy for cell sleeping and interference coordination. These works exploit medium- to long-term traffic fluctuations to consolidate load into fewer active cells through coordination.

However, as demonstrated in this and other works, small-timescale cell sleeping opportunities exist to save energy in individual cells without requiring cell coordination. As explained in \S\ref{sec:background}, Discontinuous Transmissions (DTX)~\cite{3GPP_TS_38_321} allow the transmitter to deactivate hardware components (e.g., power amplifier in the RU) during idle periods with minimal overhead, conserving power. DTX has also been explored in the literature; for example, \cite{dtx_sleeping} derives a hysteresis-based sleeping strategy considering the maximum allowable delay.

While ASMs~\cite{oran-asm} offer the potential for greater energy savings than DTX alone, they may also increase user delay due to transmissions potentially being deferred until the BS is fully operational.
Several studies have proposed strategies to optimize ASM usage \cite{salem2019traffic, asm_digital_twin, asm_distributed_q_learning, asm_multi_agent, asm_q_learning, asm_tradeoff_energy_delay, salem2019optimal}. However, these previous works have significant limitations, hindering their integration into real-world systems:

\noindent\emph{Suboptimal ASM scheduling strategies.} There is no one-size-fits-all solution that can accommodate the diverse range of real-world bursty traffic conditions. For instance, some approaches, such as those in \cite{salem2019traffic, asm_multi_agent}, propose transitioning from the deepest sleep mode (ASM 3) through intermediate sleep modes (ASM 2, ASM 1) before waking up during an idle cycle. In contrast, others like \cite{asm_tradeoff_energy_delay} suggest starting with ASM 1 and progressively entering deeper modes until ASM 3 is reached before waking up. Both strategies employ elaborate schemes (e.g., data-driven models) to optimize the duration of each sleep mode, aiming to balance energy savings against the extra delay caused by mode transitions. However, these strategies can be suboptimal under realistic bursty traffic conditions: the former may overuse deep sleep modes in high or medium-load periods, while the latter may do so under low load conditions. As demonstrated in this paper, simpler ASM scheduling strategies provide similar performance.

\noindent\emph{Unrealistic assumptions.} Many of these works rely on unrealistic assumptions, such as Poisson or log-normal traffic arrivals~\cite{salem2019traffic, asm_distributed_q_learning, asm_multi_agent}. While these models are suitable for long-term traffic behavior, they fail to capture the bursty nature of traffic at millisecond-level granularity (see \S\ref{sec:exp_analysis}). Therefore, it is key to validate ASM strategies using realistic traffic models or, as in our paper, traces collected from real base stations.

\noindent\emph{Lack of hard QoS guarantees.} Clearly, ASM utilization entails a trade-off between energy savings and delay. However, most related works do not consider hard QoS constraints, opting instead to approximate the constrained problem using Lagrange-like utility functions suitable for data-driven models \cite{salem2019traffic, asm_digital_twin, asm_distributed_q_learning, asm_multi_agent, salem2019optimal}. While acknowledging this issue, \cite{asm_digital_twin} proposes a risk model to mitigate scenarios where traffic exists but the BS is sleeping. However, varying delay requirements across different applications, services, or network slices motivate sacrificing delay for further energy savings, provided these requirements are met. This necessitates a solution with hard delay guarantees, which, to our knowledge, our paper is the first to provide in this context.

\noindent\emph{Compliance with real system constraints.} All the previous works propose performing ASM scheduling, which is a real-time (sub-ms) operation in the BS, based on rather complex methods (most of which based on reinforcement learning models)~\cite{salem2019traffic, asm_digital_twin, asm_distributed_q_learning, asm_multi_agent, asm_q_learning, asm_tradeoff_energy_delay, salem2019optimal}, which are hard to implement under such computing constraints~\cite{gpf}. As we showed in this paper, simpler ASM scheduling policies, suitable for real-time operation in the BS, are sufficient as long as joint ASM/radio policies optimized in near-real-timescales are in place, which is what \name achieves.

%% file: conclusions.tex
\section{Conclusions}

In contrast to related work that proposes complex ASM optimization algorithms for energy savings in the Radio Unit (RU) of 5G base stations, this paper introduces \name{}, a novel approach leveraging simple, real-time ASM and radio scheduling policies jointly optimized in near-real time by an O-RAN xApp. \name's unique strength lies in a novel ML framework that dynamically adapts to varying network conditions, such as the addition or removal of network slices, while maintaining pre-defined QoS requirements. Through rigorous evaluation using real-world traffic patterns and a commercial RU, we demonstrate \name's ability to achieve substantial energy reductions between 15\% and 72\% in the RU while preserving QoS constraints, establishing it as a practical solution for cost- and energy-efficient 5G networks.

%% file: ack.tex
\section{Acknowledgments}
Work supported by the European Commission through grants No. 101139270 (ORIGAMI) and SNS-JU-101097083 (BeGREEN), by NextGeneration EU through UNICO I+D grant no. TSI-063000-2021 (OPEN6G), and by the CERCA Programme.
\vspace{-2mm}